\documentclass[journal = nalefd,manuscript=article,biochem]{achemso}
\setkeys{acs}{articletitle = true}
\usepackage[utf8]{inputenc}
\usepackage{gensymb}
\usepackage{chemformula}
\usepackage{graphicx}
\graphicspath{ {./figures/} }

\author{F\'elix Thouin}
\affiliation{Department of Engineering Physics, \'Ecole Polytechnique de Montr\'eal, Montr\'eal, H3T 1J4, QC, Canada}
\author{David M. Myers$^*$}
\affiliation{Department of Engineering Physics, \'Ecole Polytechnique de Montr\'eal, Montr\'eal, H3T 1J4, QC, Canada}
\author{Ashutosh Patri$^*$}
\affiliation{Department of Electrical Engineering , \'Ecole Polytechnique de Montr\'eal, Montr\'eal, H3T 1J4, QC, Canada}
\author{Bill Baloukas}
\affiliation{Department of Engineering Physics, \'Ecole Polytechnique de Montr\'eal, Montr\'eal, H3T 1J4, QC, Canada}
\author{Ludvik Martinu}
\affiliation{Department of Engineering Physics, \'Ecole Polytechnique de Montr\'eal, Montr\'eal, H3T 1J4, QC, Canada}
\author{St\'ephane K\'ena-Cohen}
\email{s.kena-cohen@polymtl.ca}
\affiliation{Department of Engineering Physics, \'Ecole Polytechnique de Montr\'eal, Montr\'eal, H3T 1J4, QC, Canada}

\title{Broadband field-enhancement in epsilon-near-zero photonic gap antennas}
\date{December 2021}
\SectionNumbersOff
\abbreviations{ENZ,PGA,ITO,THG,DFSS}
\begin{document}
\maketitle
\begin{abstract}
In recent years, the large electric field enhancement and tight spatial confinement supported by the so-called epsilon near-zero (ENZ) mode has attracted significant attention for the realization of efficient nonlinear optical devices. Here, we experimentally demonstrate a new type of antenna, termed an ENZ photonic gap antenna (PGA), which consists of a dielectric pillar within which a thin slab of indium tin oxide (ITO) material is embedded. In ENZ PGAs, hybrid dielectric-ENZ modes emerge from strong coupling between the dielectric antenna modes and the ENZ bulk plasmon resonance. These hybrid modes efficiently couple to free space and allow for large enhancements of the incident electric field over nearly an octave bandwidth, without the stringent lateral nanofabrication requirements required by conventional plasmonic or dielectric nanoantennas.  The linear response of single ENZ PGAs is probed using dark field scattering and interpreted using a simple coupled oscillator framework. Third harmonic generation is used to probe the field enhancement and large enhancements are observed in the THG efficiency over a broad spectral range. This proof of concept demonstrates the potential of ENZ PGAs, which we have previously shown can support electric field enhancements of up to 100--200X, which is comparable with those of the best plasmonic antennas.
\end{abstract}

\section{Introduction}

Advances in optical signal processing, optical neural networks and photonic quantum computing are urgently driving the need for nonlinear optical components with ever smaller footprints and lower energy consumption. In this context, epsilon-near-zero (ENZ) materials have attracted significant attention. In an ENZ material, the real part of the permittivity crosses zero due to the presence of a natural (e.g. plasma, phonon, exciton) or artificial resonance (e.g. metamaterial).  This results in exotic optical phenomena such as infinite phase velocity\cite{edwards_experimental_2008} and vanishing photonic density of states\cite{lobet_fundamental_2020}.  Notably, thin films of epsilon near-zero (ENZ) materials have been reported to have exceptionally high optical nonlinearities\cite{passler_second_2019,reshef_nonlinear_2019,wen_doubly_2018}. Doped semiconductors are particularly interesting as a class of ENZ materials, given that their plasma frequency can be readily controlled via the dopant concentration and tuned in-situ through electrostatic gating. To a large extent, however, the nonlinearities reported in ENZ materials are a consequence of the field enhancement intrinsically provided by ENZ thin films \cite{vassant_berreman_2012,campione_theory_2015}. Importantly, ENZ films support so-called Berreman modes, which can be directly excited within the light cone using TM-polarized light. These lead to modest enhancements of the incident electric field for high angles of incidence that can nevertheless strongly enhance high-order nonlinear processes \cite{alam_large_2018}. For a thin indium tin oxide (ITO) layer serving as the ENZ medium, for example, a typical enhancement is $E_{ITO}/E_0 \sim 2-3$ can be anticipated\cite{campione_epsilon-near-zero_2015,alam_large_2018}, where $E_0$ and $E_{ITO}$ are the magnitudes of the electric field in air and ITO, respectively. In addition, ultrathin ENZ films support a so-called ENZ mode, which is nearly dispersionless and allows for deep sub-wavelength confinement\cite{campione_theory_2015}. This mode can show much larger field enhancements (typ. $E_{ITO}/E_0 \sim 10-20$), but it exists beyond the light cone and can only be excited using e.g. a Kretschmann configuration (ideally at the critical coupling angle) or some form of mode matching with free space.\cite{luk_enhanced_2015}. In both cases, enhancements typically occur over a 100--200 nm bandwidth around the ENZ frequency. Many experiments have explored such effects by placing either metal \cite{alam_large_2018,hendrickson_coupling_2018,kim_role_2016,schulz_optical_2016,campione_near-infrared_2016,jun_epsilon-near-zero_2013} or dielectric \cite{wang_large_2022} optical antennas on ENZ thin films to increase coupling with free space, provide further lateral confinement or engineer the spectral response. 

We have previously reported that all-dielectric photonic gap antennas (PGAs) consisting of a thin low-index medium, sandwiched in a high index dielectric pillar, can provide extremely large Purcell factors, strong directionality and large field enhancements over a broad spectral range\cite{patri_photonic_2021}. These rely principally on the electric field enhancement ensured by the continuity of the displacement field across the interface between high ($\epsilon_{high}$) and low permittivity ($\epsilon_{low}$) media $\epsilon_{low}E_{low}=\epsilon_{high}E_{high}$, where $E_{low/high}$ are the longitudinal components of the electric field near the interface. Photonic gap antennas also have the significant advantage of not requiring any deep sub-wavelength lithography as is typically required for most optical nanoantennas. They instead simply rely on the fabrication of a very thin low-index layer. In the extreme case, where the thin layer is an ENZ material $\epsilon_{low}\sim 0$, we have shown that the bulk plasmon mode hybridizes with the bare modes of the all-dielectric antenna, leading to hybrid modes with extremely large field enhancements.\cite{patri_hybrid_2022} Full wave electromagnetic calculations show that such structures can provide field enhancements of $\sim 100$ using 2 nm-thick films of ITO and $\sim 150$ in the mid-infrared using 20 nm-thick doped GaAs, as the ENZ media. So far, ENZ PGAs have not been experimentally realized. In this letter, we report on the design, fabrication and characterisation of dielectric PGAs hosting a 10 nm-thick layer of ITO as the ENZ material. Using dark field scattering spectroscopy, we observe several resonances in the infrared. Using frequency domain finite-element method simulations of the far-field scattering and field enhancement spectra, these can be assigned to the different polaritonic modes supported by the PGA.  We then probe the electric field enhancement inside the ENZ layer by measuring third harmonic generation (THG).
This also reveals strong resonances in the third harmonic generation efficiency at the energy of the hybrid modes, evidence of the strong subwavelength electric field confinement.
Peak instantaneous THG efficiencies of about 100 $\mu$m$^4$/MW$^2$ are obtained at these resonances, with a strong response spanning $>500$~nm.
Despite strong losses in the antenna's dielectric, these efficiencies exceed those recently reported of for mutliresonnant Ge nanodisk antennas\cite{grinblat_degenerate_2017} and could be significantly improved through the use of thinner ITO layers.\cite{Datta2020}

\section{Results and discussion}

\begin{figure}
    \centering
    \includegraphics{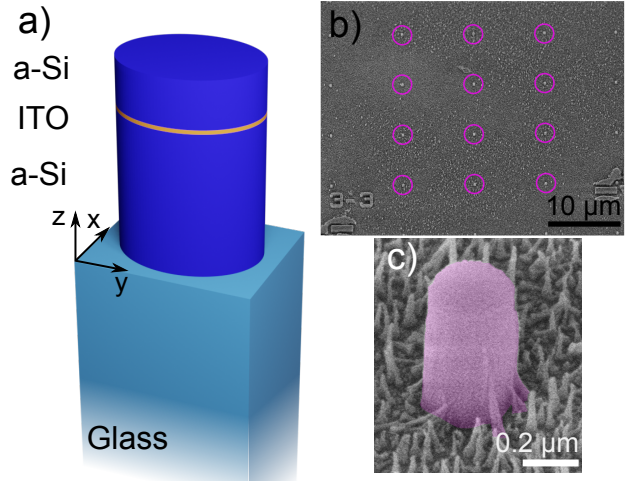}
    \caption{Design and dimensions of the photonic gap antennas. a) A render of the PGA design. It consists of amorphous silicon pillars of elliptical cross section in which a thin ITO layer is embedded. The antennas stand on a thick glass substrate. The thickness of the glass/a-Si/ITO/a-Si layers are 2\;mm, 373\;nm, 10\;nm and 128\;nm respectively. b) SEM image of an antenna array on the sample. The antennas are encircled. c) A close-up on a single antenna, highlighted for clarity. The fabricated antennas are faithful to the design despite thin grass-like structures around them and slightly bent sidewalls.}
    \label{fig:sem}
\end{figure}

A schematic of the fabricated PGA is depicted in Figure \ref{fig:sem} a).
It consists of a high refractive index pillar of elliptical horizontal cross section in which a thin layer of ENZ material is embedded.
The design stems from considerations discussed in a previous publication \cite{patri_photonic_2021}. 
The ENZ layer is placed at three quarters of the pillar's height to improve the directivity of the antenna through modal interference.
The pillar's elliptical cross section is chosen to improve the antenna's field enhancement by concentrating the light intensity midway through the pillar's major radius. For operation in the near infrared, amorphous silicon (a-Si) and ITO are chosen as the high refractive index and ENZ materials respectively. Indium tin oxide, which is a conductive oxide commonly used as a transparent conductor, has an ENZ frequency in the near infrared. The exact crossing point of the real part of the permittivity, which is a consequence of the plasma frequency, and thus the carrier concentration, can be strongly tuned through the oxygen vacancy density\cite{cleary_optical_2018}.
In our highly conductive samples, the ITO ENZ wavelength was measured to be 1188\;nm using variable angle spectroscopic ellipsometry (see Supplementary Figure S1) performed on a 30 nm-thick film grown under the same conditions as those used in our PGA. 

The fabricated elliptical a-Si/ITO/a-Si PGAs have layer thicknesses of 373\;nm, 10\;nm and 128\;nm, respectively. While the predicted field enhancement could be further increased for even thinner ITO films, 10 nm was chosen as it allows for relatively uniform films to be achieved directly using reactive sputtering under a wide range of conditions (see Supplementary Information Figure S2). Achieving high quality thinner (down to monolayer) films is possible, but requires substantial optimization of the deposition conditions or specialized printing techniques.\cite{Li2019,Datta2020} Antennas were patterned laterally using electron beam lithography (see Methods). Each fabricated array holds nominally identical pillars spaced 10\;$\mu$m apart to avoid coupling between pillars.
To investigate the impact of the antenna dimensions on the response, three arrays (I, II and III) hosting antennas with different cross-sectional dimensions were investigated.
Scanning electron microscope (SEM) micrographs of one of these arrays as well as a close up on a single antenna are shown in Figures \ref{fig:sem} b) and c) respectively.
The cross-sectional dimensions of the antennas in these arrays are summarized in Table\;\ref{tab:pillardimensions}.
These were measured using top-down SEM micrographs shown in Figure S3 of supplementary information.
These micrographs confirm that the fabricated structures are faithful to the design, but with some minor imperfections such as a slightly narrower top and the presence of micrograss between the antennas.
The latter occurs to due to the random formation of silicon oxyfluoride micromasks during the etching of the bottom a-Si layer.\cite{jansen_black_1995}
However, as we will demonstrate with further optical characterisation, these have no measurable impact on the antenna response.

\begin{table}
    \centering
    \begin{tabular}{c|c|c}
    Array index & Minor diameter (nm) & Major diameter (nm)  \\
    \hline
    I&$230\pm10$&$290\pm10$\\
II&$265\pm5$&$320\pm7$\\
III&$290\pm10$&$365\pm5$\\ 
    \end{tabular}
    \caption{Cross-sectional dimensions of the investigated antennas}
    \label{tab:pillardimensions}
\end{table}

\begin{figure}
    \centering
    \includegraphics{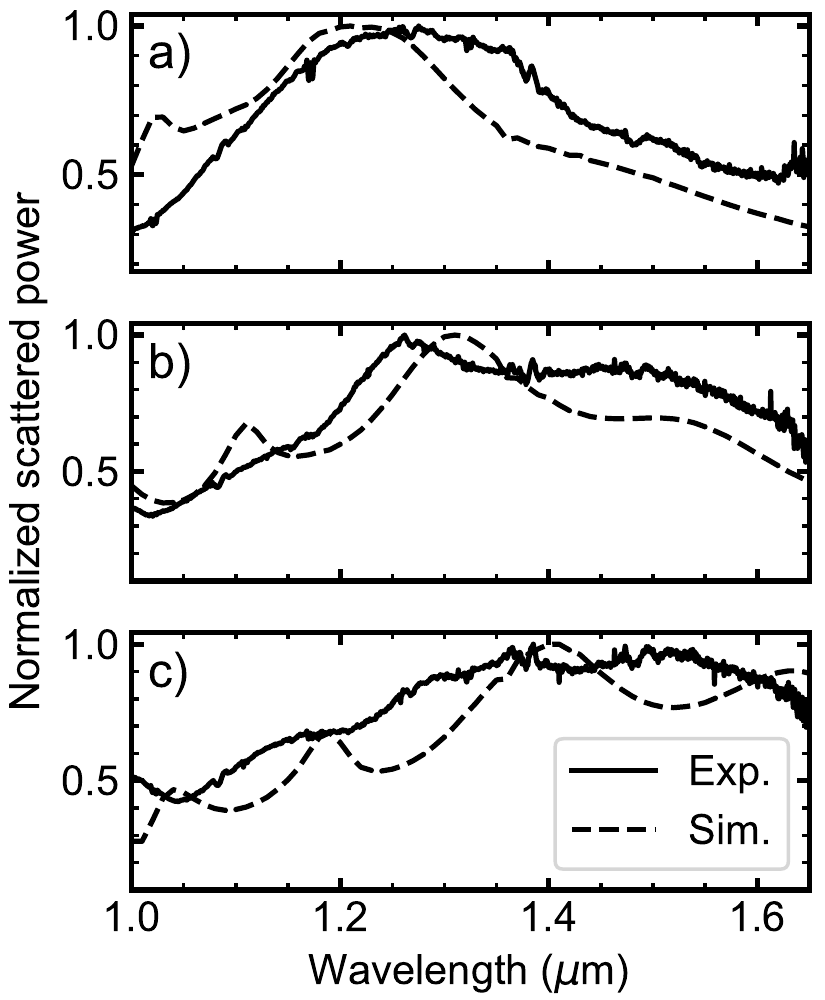}
    \caption{Experimental (solid lines) and simulated (dashed lines) dark field scattering spectra  of a single antenna in arrays I, II and III in a), b) and c) respectively.}
    \label{fig:dfss}
\end{figure}

To probe the linear optical response of the antennas, we first measure the far-field scattering spectrum of individual antennas in arrays I, II and III using single particle dark field scattering (DFS) spectroscopy. 
The DFS spectra for an antenna representative of those in arrays I, II and III are plotted in Figure \ref{fig:dfss} a), b) and c) respectively.
The DFS spectra from other antennas of the same array are qualitatively similar with slight variations in peak intensities and position due to structural inhomogeneities between nominally identical antennas (see additional DFS spectra in Figure S3 of supplementary information).
The micrograss surrounding the antennas do not lead to any observable dark field scattering in the spectral region probed by our experiment (see Figure S4 of supplementary information).
The broadband DFS response covers nearly an octave and redshifts as the lateral dimensions of the antennas are increased from array I to III.\\

In Figure \ref{fig:dfss}, we also compare our experimental data with the simulated scattering cross-section integrated over the collection objective's numerical aperture.
The dimensions reported in Table \ref{tab:pillardimensions} were used for these calculations. 
The result for arrays I, II and III are plotted as dashed lines alongside the experimental data in Figure \ref{fig:dfss} a), b) and c) respectively.
The resonant structure observed in the DFS spectra is faithfully reproduced for all dimensions but with narrower, slightly shifted features.
These discrepancies are most likely due to small deviations from the idealized geometry and increased damping for the 10\;nm film as compared to the measured refractive index.

\begin{figure}
    \centering
    \includegraphics{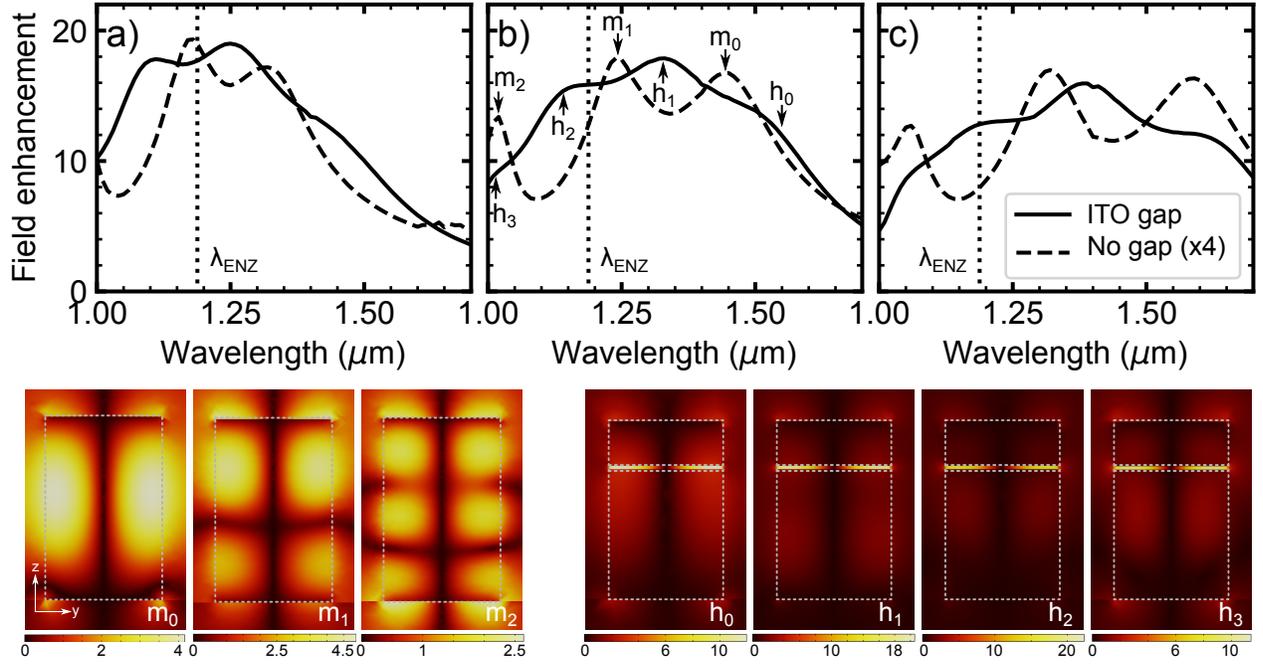}
    \caption{Simulations of the field enhancement for the antennas in arrays I, II and III (in a), b) and c) respectively) with (full lines) or without (dashed lines) an ITO gap. The dotted line indicates the ENZ wavelength. Cuts in the electric field profile's z-component under normal excitation are shown for the labelled modes in their respective panels. These cuts are taken in the major axis plane of antenna II. The dotted lines outline the boundaries of the antenna. The color map gives the local field enhancement $|E_z|/|E_0|$}
    \label{fig:fieldenh}
\end{figure}

The agreement between the experimental DFS spectra and the simulated ones allows us to interpret the behavior of the antennas using the quasinormal modes of the simulated structures.
The DFS spectra capture the far-field properties of the antennas.
The near-field properties on the other hand, are better captured by the field enhancement spectra plotted in Figure \ref{fig:fieldenh}.
We define the field enhancement as the maximal ratio between the z-component of the electric field in the antenna to the magnitude of the incident electric field.
The figure also shows the case of ungapped antennas to highlight the effect of the interaction between the ITO slab and the bare dielectric antenna modes.
The antennas show very distinct resonances within the spectral region covered by the DFS bands.
Like the DFS spectra, when the cross-sectionnal dimensions of the antennas are increased from Figure \ref{fig:fieldenh} a) to c), these resonances redshift and spread apart.
Compared to the purely dielectric antennas, the ENZ PGA field enhancement spectra exhibit broader, shifted features, an additional resonance, and a fourfold increase in field enhancement.
This behavior is a sign of the strong coupling between the dielectric antenna modes and the ITO plasmonic resonance at the ENZ wavelength. \cite{patri_hybrid_2022}
While the interpretation is the same for all antenna dimensions, the modes are labelled and shown for the case of antenna II.
Dielectric modes are identified by resonances in the ungapped antenna's field enhancement spectra such as those labelled m$_0$, m$_1$ and m$_2$ in Figure \ref{fig:fieldenh} b).
A cross-section of the z-component of their internal electric field reveals horizontally anti-symmetric field distributions for all modes and an increasing number of vertical nodes with the mode index.
The strength of the interaction between the bulk plasmon resonance and a given dielectric mode is related to the electric field's z-component.
By placing the ENZ layer at a quarter of the antenna's height from the top, the ENZ mode couples most strongly to mode m$_1$ then m$_0$ and least to m$_2$.
This coupling creates new modes labelled h$_0$, h$_1$, h$_2$ and h$_3$ in Figure \ref{fig:fieldenh} b).
These are shifted from the bare antenna modes by their interaction with the ENZ mode and broadened due to its increased damping.
These hybrid light-matter modes inherit both the bare antenna modes' coupling to the far-field and the strongly localized character of the ITO plasmonic resonance.
These characteristics are clearly manifested in the antenna's electric field profiles plotted in the mode-labelled panels of Figure \ref{fig:fieldenh} for the antennas of array II. 
All profiles taken within the field enhancement band show both a strong localization of the electric field within the ITO layer and a radiative component in and around the antenna.
This ultimately manifests in the broadband field enhancements of 10-20 shown in Figure \ref{fig:fieldenh} as well as the efficient coupling to far-field radiation.
The same interpretation holds for antennas I and III.
It must be noted however, that some modes do not couple strongly to the ENZ layer.
This is the case for mode h$_3$ of antenna II, for example.
Here, the ITO slab is located close to the node of the photonic mode m$_2$, leading to a weak coupling with the ENZ mode.

\begin{figure}
    \centering
    \includegraphics{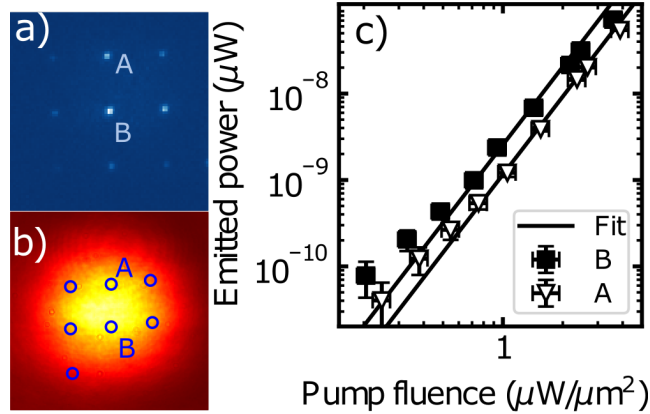}
    \caption{a) Third harmonic image of pillar array II illuminated by a pulsed IR pump at 1.3\;$\mu$m.
    Antennas are identified through their bright emission.
    Two of them are labelled A and B.
    b) The IR pump profile at the sample plane used to measure the pump fluence at the location of each pillar.
    Circles are positioned at the corresponding pillar locations extracted from the third harmonic image in a).
    The location of antennas A and B is labelled accordingly.
    c) Third harmonic power emitted from A (triangles) and B (squares) for different pump fluences and their fits to a cubic dependance on pump fluence (solid lines).
    Vertical error bars represent the contribution of shot noise and background noise while horizontal error bars represent uncertainties on the pump power during the measurement.}
    \label{fig:thgpower}
\end{figure}

We next probe the field enhancement occurring within the fabricated devices through THG.
As any non-linear optical phenomenon, THG is a highly sensitive probe of the local field intensity created by a pump beam. As such, it has been widely used to probe field enhancements in metallic\cite{hanke_efficient_2009} and dielectric\cite{xu_boosting_2018} optical nanoantennas, thin films of ENZ materials\cite{capretti_enhanced_2015,luk_enhanced_2015} and metasurfaces\cite{tong_enhanced_2016}.
Our experimental setup allows us to image emission from the sample at the pump's third harmonic and resolve the contribution of individual pillars (see Methods section for more details).
The image of pillar array II illuminated by a 1.3\;$\mu$m pump, taken at the THG frequency, is shown in Figure\;\ref{fig:thgpower} a).
Intense emission is clearly visible at the antenna positions.
No signal above the background is observed between the antennas, confirming the large field enhancement and the negligible role played by the grass-like background.
To confirm that this emission stems indeed from THG, we check that it follows a cubic dependence on pump fluence.
We compensate for variation in the pump profile over an array by acquiring an image of the pump spot at the sample such as the one shown in Figure\;\ref{fig:thgpower} b) (see Methods for details).
The power emitted from two antennas (A and B in Figure\;\ref{fig:thgpower}) are plotted against pump fluence in Figure\;\ref{fig:thgpower} c).
Both datasets are well fitted over three ordes of magnitude by a cubic dependance on pump fluence, confirming that the observed emission stems from THG inside the antenna.
We cannot discard, however, the contribution of cascaded sum frequency processes to this emission\cite{celebrano_evidence_2019}.
Nevertheless, cascaded or not, THG remains a sensitive probe of field enhancement inside the antennas.

Despite being nominally identical, pillar B is about twice as brigh as pillar A. This reflects the presence of important structural differences between the antennas to which neither SEM nor DFS spectroscopy are sensitive to.
This hints at the location of the THG process within the antennas.
Great structural differences between antennas are unlikely to arise from the bulk of the silicon layers due to their amorphous structure, or from the antennas' surfaces given their smooth profiles (see Figure S2b of supplementary information).
On the other hand, the polycrystalline ITO layer is noticeably rough and features clear crystalline domains of tens of nanometers in size (see Figure S2a of supplementary information).
When all but the antennas' cross sections are etched away from the a-Si/ITO/a-Si multilayer, each antenna samples a different region of the inhomogeneous ITO layer.
Since the antenna's radiative properties depend on the overall mode profile, the DFS spectra are only weakly affected by variations in the microstructure.
However, THG is highly sensitive to the permittivity at the mode hotspot.
Since the hotspot is concentrated in the ITO, as shown in the panels of Figure\;\ref{fig:fieldenh}, the microstructure variations are likely to cause the THG intensity variation between antennas.

We now check the contributions of the different antenna modes to THG. 
We can extract the antenna's TH power conversion efficiency by dividing the time averaged emitted TH irradiance $\langle P_{\textrm{THG}}\rangle$ by the time-averaged received pump irradiance $\langle P_{\textrm{pump}}\rangle $.
Given that this number depends on $\langle P_{\textrm{pump}}\rangle $ and the pump pulse duration, a more meaningful result is obtained by instead defining the THG efficiency as the ratio of the peak TH irradiance to the cube of the peak pump irradiance.
The observed cubic dependance on pump irradiance guarantees this metric to be independant of pump irradiance while using peak irradiance values accounts for the pump pulse duration.
Lastly, the values reported here strongly underestimate the internal THG efficiency due to strong absorption of the TH within the a-Si dielectric and the finite numerical aperture of the collection objective. Simulations shown in Supplementary Information estimate that 98\% of the generated TH power is absorbed by the a-Si and only 23\% of the radiated TH is collected by our apparatus due to the highly anisotropic THG radiation profile.

\begin{figure}
    \centering
    \includegraphics{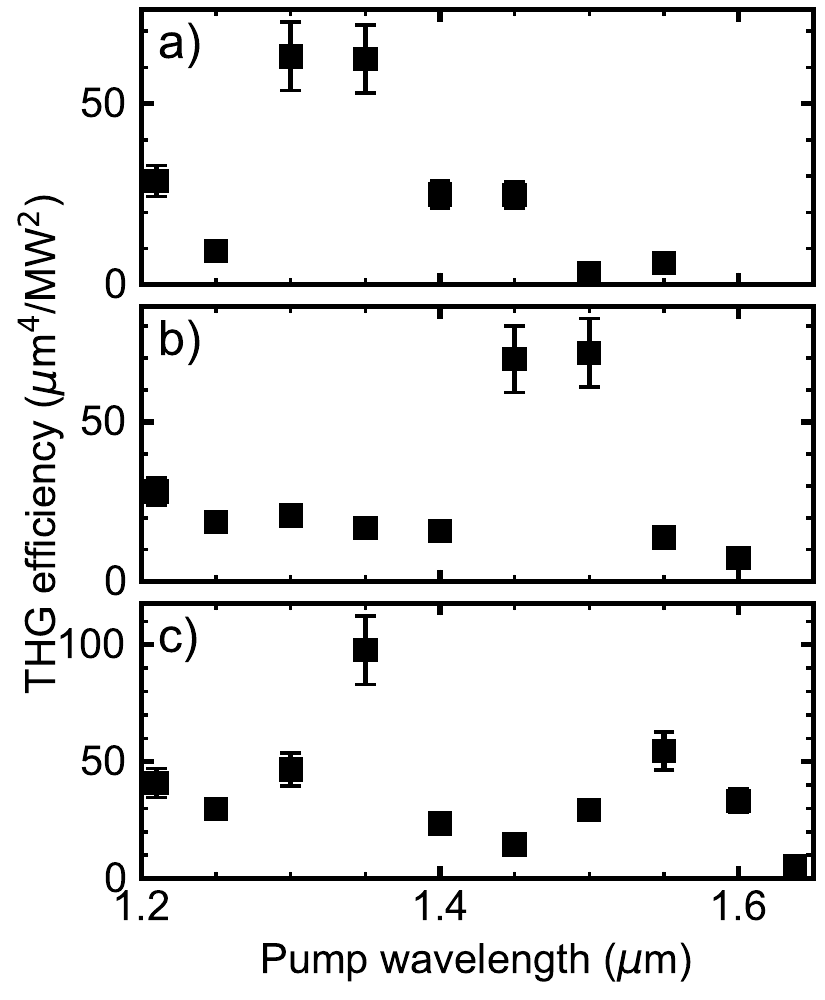}
    \caption{Pump wavelength dependence of the measured instantaneous THG efficiency for an antenna in arrays I (a), II (b) and III (c) (black squares). Error bars represent the effect of compounded uncertainties in the calculated values.
    All curves correspond to the same antennas investigated in Figure\;\ref{fig:dfss}.}
    \label{fig:THGWave}
\end{figure}
We measured the instantaneous THG efficiency of antennas within arrays I, II and III using pump wavelengths spanning 1.21\;$\mu$m to 1.64\;$\mu$m and plotted them in Figure\;\ref{fig:THGWave} a) to c).
The single antennas chosen for this investigation are the same as those whose DFS spectra are plotted in Figure \ref{fig:dfss}. For all antennas, $\sim100$~nm wide resonances are observed atop a broad THG response within the measured wavelengths.
Just like in the DFS spectra, these resonances redshift with increasing lateral pillar dimensions.
They also consistently match the experimental peaks observed in the DFS spectra of Figure \ref{fig:dfss}.
However, despite showing strong THG over their whole field enhancement band, the shape of the THG efficiency curves differ from the calculated field enhancement spectra cubed. Notably, the highest enhancement is observed for the widest antenna, in contrast to the prediction of Fig.~\ref{fig:fieldenh}. Several mechanisms can contribute to these differences. In particular, the a-Si's absorption nearly triples over the probed TH spectral window of 550\;nm to 400\;nm\cite{pierce_electronic_1972}, which significantly reduces the radiated THG power at shorter wavelengths. In addition, the longitudinal and transverse components of ITO's frequency dependent nonlinear susceptibility have been reported to increase by approximately $60\%$ and $100\%$, respectively, in the range between 1020\;nm to 1320\;nm\cite{rodriguez-sune_retrieving_2021}. Finally, nonlocal effects, which have been ignored, and material roughness can  play an important role in the generation and coupling of localized THG to far field radiation.
While all antennas show strong THG over the spectral region probed, the amplitude of the resonances vary significantly between pillars within the same array (see supplementary material).
This reflects the important local inhomogeneities between nominally similar antennas previously highlighted.
For comparison's sake, TH power conversion efficiencies measured for the antennas studied in Figure\;\ref{fig:THGWave} are 8.6$\times 10^{-8}$ at 3.7 GW/cm$^2$, 4.7$\times 10^{-8}$ at 2.6 GW/cm$^2$ and 5.1$\times 10^{-8}$ at 3.2 GW/cm$^2$ for arrays I, II and III excited at 1.35, 1.45 and 1.55 $\mu$m respectively. These figures include the large absorption losses in the a-Si and the limited collection efficiency. From simulations of the THG external coupling efficiency, we expect internal efficiencies to be 200 times larger than the figures reported here. Given that the dominant loss mechanism is simply re-absorption of the THG radiation within the a-Si, moving to a longer wavelengths or a transparent dielectric could alleviate these losses.

\section{Conclusion}

We have demonstrated photonic gap antennas in which an ultra thin layer of ITO, an ENZ material, is embedded.
By comparing dark field scattering spectroscopy measurements with simulations of the antennas' scattering cross-section, we showed the antennas host hybrid polaritonic modes in accordance with simulations and previous work. Through third harmonic generation, we confirm that these modes tightly confine the fields inside the 10\;nm thick ITO layer, leading to high field enhancement over a broad spectral range of 400\;nm. This proof of concept demonstrates the effectiveness of PGAs as vertically stacked optical antennas.
They afford a strong localized field enhancement in an device that can be fabricated using photolithography or straightforward electron-beam lithography, forgoing the high lateral resolution requirements of traditional antenna designs.
While our demonstration is in the near-infrared, PGAs are particularly attractive for applications in the mid-infrared where the ENZ frequency of doped III-V semiconductors is located\cite{jun_doping-tunable_2014} .
Using molecular beam epitaxy, infrared ENZ materials could be grown inside PGAs with deep subwavelength thicknesses well below those reported here and provide more than an order of magnitude larger field enhancement.

\section{Acknowledgements}
This work was supported by the Natural Sciences and Engineering Council of Canada Strategic Grant Program, the Canada Research Chairs Program and the Discovery Grants Program. We would also like to acknowledge design and fabrication support from CMC Microsystems and Canada’s National Design Network (CNDN).

\subsection{Author Contributions}
* F.T, D.M.M and A.P have contributed equally to this work.
\section{Methods}

\subsection{Sample fabrication}

A B270 glass substrate (51\;mm x 51\;mm and 2\;mm thick) was pre-cleaned using an oxygen and argon plasma.
The PGAs were fabricated by sputtering in a CMS-18 deposition system by Kurt J. Lesker using 3-inch targets of Si and ITO (In2O3/SnO2 90/10 wt \%). Each layer of material was grown on a glass substrate to the desired thicknesses.
A 372\;nm layer of amorphous silicon was then sputtered at 200\;\degree C, followed by a 10\;nm layer of ITO sputtered at room temperature.
The sample was then removed from the chamber and measured on a J.A. Woollam RC2-XI ellipsometer.
After returning it to the sputtering chamber, it was annealed under vacuum for an hour at a 400\;\degree C setpoint in order to reduce the epsilon-near-zero wavelength to 1188\;nm.
The temperature was then brought down to 200\;\degree C for the sputtering of the final 128\;nm layer of amorphous silicon. 
The layer thicknesses and bulk optical properties were measured using the same ellipsometer (see Figure S1 of supplementary information).  

The large sample was cleaved into several smaller samples roughly equal in size, which were cleaned by sonication in acetone and isopropyl alcohol followed by ten minutes in an oxygen plasma (200\;W, 0.8\;mbar).
Immediately before spin coating, each sample was vacuum baked (50\;torr, 120\;\degree C, five minutes) and primed with SurPass 3000.
They were spin-coated with MA-N 2403 (4000 rpm) negative electron beam resist, pre-baked on a hotplate (90\;\degree C, 1 minute), and exposed in a Raith 150 electron beam lithography system (20\;kV, 80\;$\mu$C/cm$^2$).
Development was done in MA-D 525 developer (40 seconds) followed by two rinses in deionized water while stirring (10 minutes total) then by a post-bake on a hotplate (90\;\degree C, 2 minutes).
The samples were then mounted onto an \ch{Al2O3}-coated silicon carrier wafer using Crystalbond and etched in an Oxford Instruments Plasmalab 100 ICP 180 reactive ion etching tool. 
The recipe for etching the top amorphous silicon (1200\;W ICP power, 60\;W RF, 49\;V bias, 20\;mtorr, 15\;\degree C, 22\;sccm \ch{SF6}, 38\;sccm \ch{C4F8}, 16\;seconds) is a pseudo-Bosch process similar to those used elsewhere for anisotropic etching of silicon \cite{con_nanofabrication_2014,saffih_fabrication_2014,khorasaninejad_enhanced_2012,ayari-kanoun_silicon_2016,henry_alumina_2009}.
The sample was removed from the chamber and wet etched in an acid solution (1:3:4 \ch{HNO3}/HCl/\ch{H2O}) for 2 minutes to remove the ITO layer.
The sample was reinserted in the chamber and etched a second time using the same peusdo-Bosch recipe for 35 more seconds, and finally cleaned in an oxygen plasma (2000\;W ICP, 0\;W RF, 35\;mtorr, 15\;\degree C, 30 seconds).

\subsection{Optical characterisation}

The optical characterisation setup used was designed to investigate both dark field scattering (DFS) and third harmonic generation (THG) in single antennas.
This allows the direct correlation of individual DFS and THG spectra despite possible inhomogeneities between pillar specimens. 
A detailed diagram of this setup is presented in the supplementary information.
The sample was held in an inverted microscope customized to allow illumination straight from the top through a 10x microscope objective for THG or at a grazing angle through a 10\;mm biconvex lens for DFS spectroscopy.
A stabilized laser lamp provided a broad and stable infrared illumination for the DFS experiments. The output of an optical parametric amplifier (200\;fs pulse duration, 100\;kHz repetition rate, Light Conversion ORPHEUS-N) was used to pump the THG.
For both experiments, light was collected from the bottom of the microscope through a 40x 0.75 NA objective and sent to an infared spectral imager for DFS scattering spectroscopy or a visible light imaging system for THG.
The collected light was directed to either systems using an accurate magnetically registered kinematic mount.
The visible imager consists of a single lens, a visible shortpass filter (400\;nm to 550\;nm bandpass, Thorlabs FESH550) and a silicon CCD camera (Princeton Instruments Pixis 400) sensitive from 350\;nm to 1000\;nm.
The infrared imager is composed of a InGaAs CCD camera (First Light C-RED2) sensitive from 800\;nm to 1650\;nm connected at the exit of an imaging spectrometer.

When performing DFS experiments, only the infrared imager was used with the spectrometer's first diffractive order. The spectrum of the broadband light source was obtained by focusing it at the sample plane from within the collection objective's numerical aperture and recording it using the infared imager.
The spectrum of the DFS cross-section was obtained by dividing the scattered spectrum by that of the light source after proper background subtraction. 

During THG measurements, the pump profile at the sample was first imaged using the infrared imager with its spectrometer at the zeroth order and a neutral density filter. The pump was then turned off with a mechanical shutter and the sample's region of interest was moved under the pump's illumination. The collection path is then reconfigured for the use of the visible imager and the pump shutter opened to acquire an image of the pump induced THG. Using a ruled sample, the infrared and visible imagers' image planes are calibrated so that a location on one of these planes can be mapped onto the other. This allows for the measurement of the pump power density responsible for THG in a given pillar. 

All pump powers were measured with a high resolution thermal power sensor (Thorlabs S401C). The third harmonic power was measured by calibrating the number of counts on the visible imager's camera to the power emitted at the sample's plane. This calibration was performed for every wavelength of interest using the same acquisition parameters as those used in the THG measurements.

\bibliography{main}
\end{document}


\maketitle
\section{Characterization of the ITO and silicon layers}
\subsection{Relative permittivity and ENZ wavelength}
\begin{figure}
    \centering
    \includegraphics[width=\textwidth]{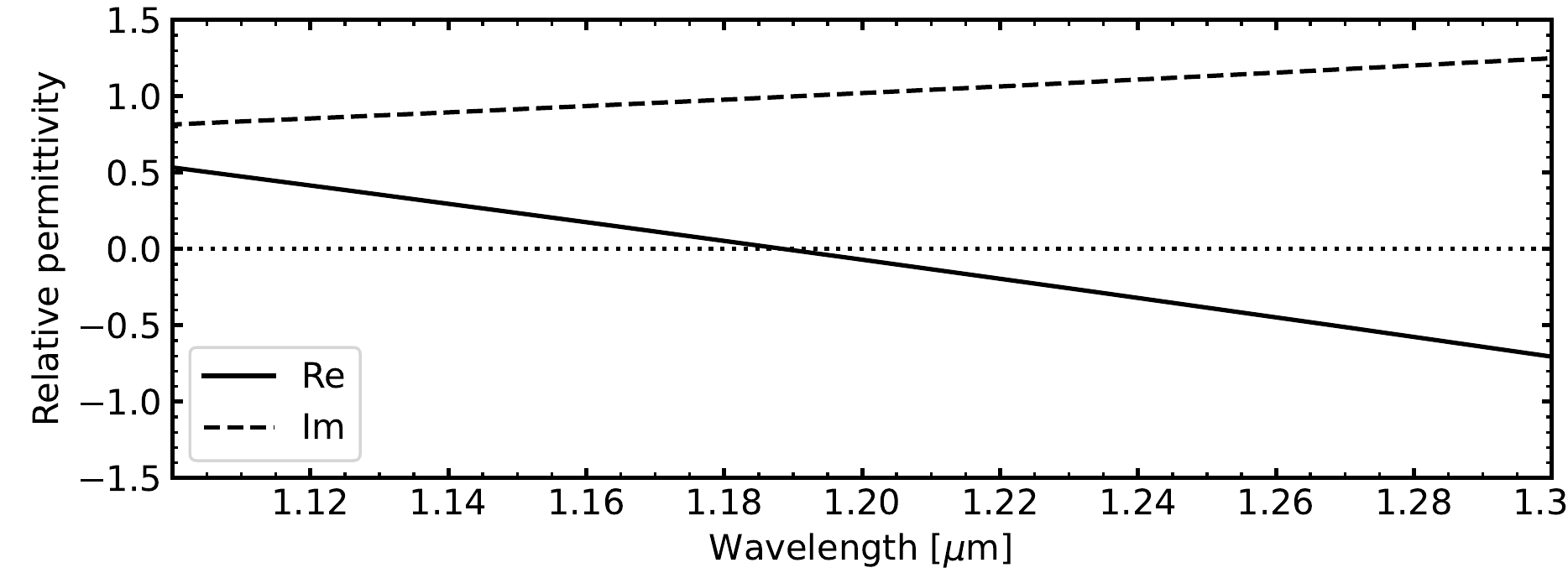}
    \caption{The real (solid) and imaginary (dashed) parts of the relative permittivity of the ITO layer used in the fabrication of the PGAs. The real part crosses zero at a wavelength of 1.188\;$\mu$m. The dotted line denotes zero permittivity.}
    \label{fig:s1}
\end{figure}

The relative permittivity of the ITO layer embedded inside the PGAs is plotted in Figure \ref{fig:s1}. From this measurement, the ENZ wavelength is determined to be 1.188\;$\mu$m. As described in the Methods section of the main text, the relative permittivity of the ITO layer was measured in situ using a M-2000 ellipsometer from J.A. Woollam. The thickness of the ITO layer was also extracted from the same measurement.
\subsection{Film morphologies}
\begin{figure}
    \centering
    \includegraphics{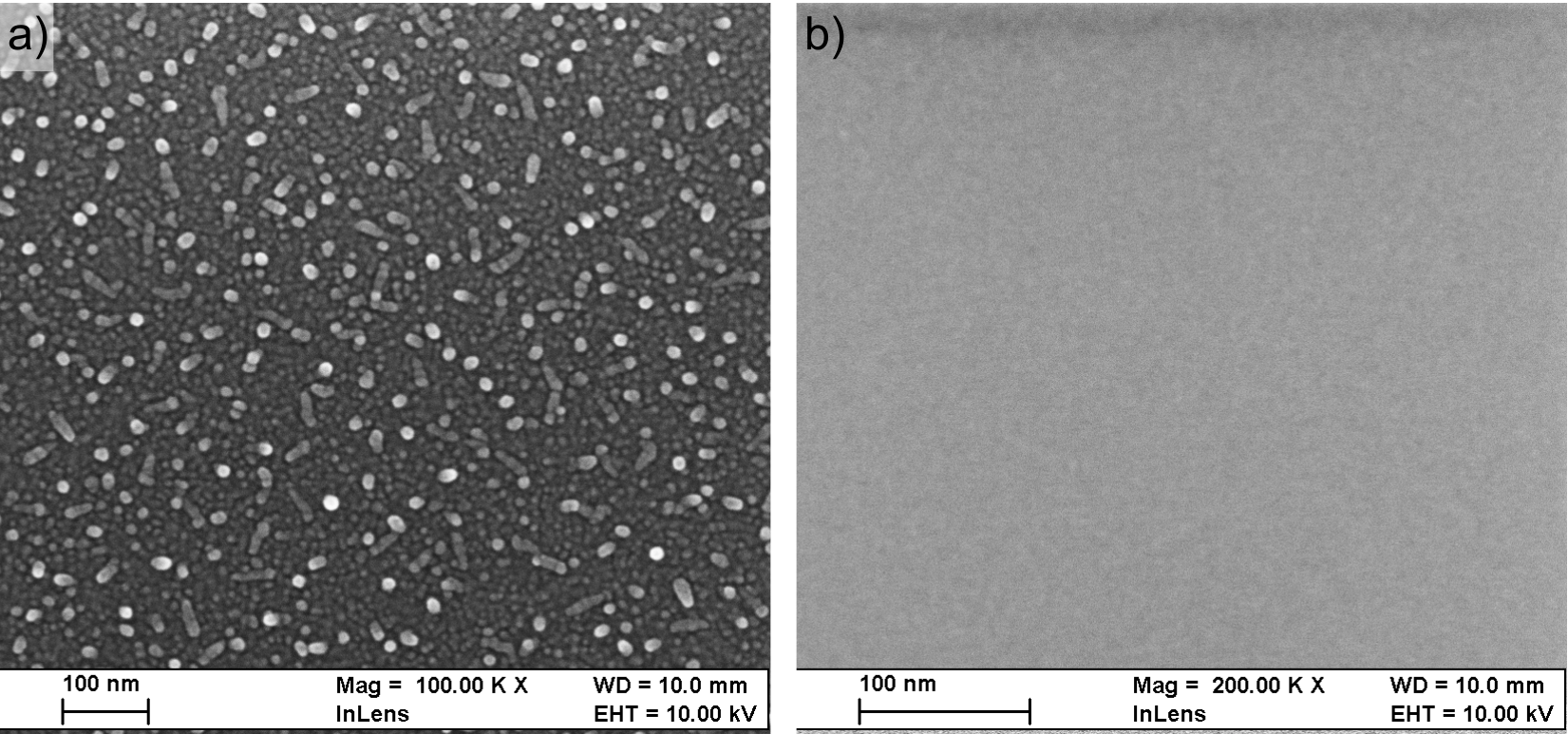}
    \caption{SEM micrograph of a) a 13\;nm thick ITO film deposited on 373\;nm of amorphous silicon and b) the surface of the amorphous silicon layer on which it was deposited.}
    \label{fig:itosem}
\end{figure}

SEM micrographs of films grown in similar conditions as those used to fabricate the antennas are presented in Figures \ref{fig:itosem} a and b for the ITO and amorphous silicon respectively. The surface of the 373\;nm thick a-Si layer is smooth and uniform while clear crystalline domains and structure is observed on the 13\;nm thick ITO film.
\section{Pillar dimensions}
\begin{figure}[h!]
    \centering
    \includegraphics[width=0.5\textwidth]{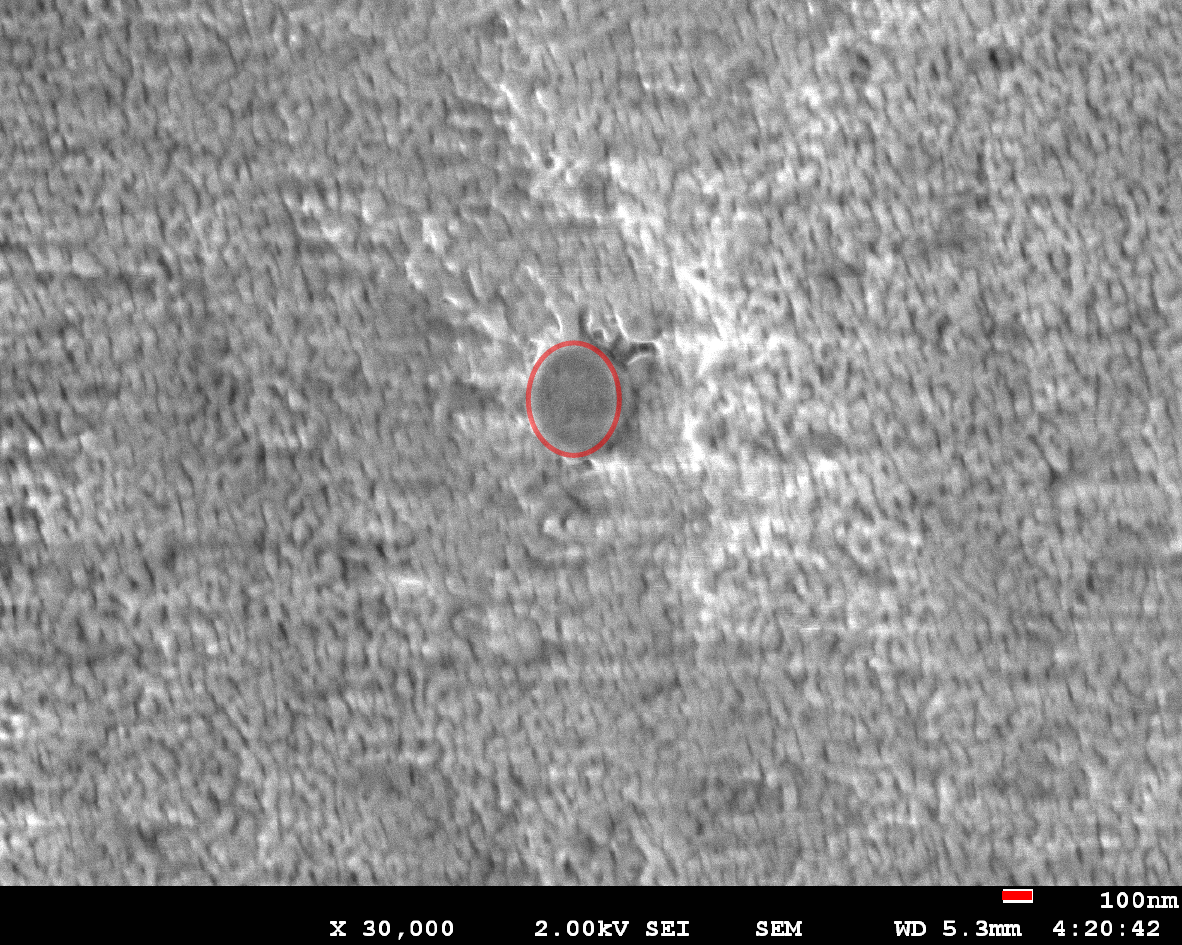}
    \caption{Top down SEM micrograph of a single pillar to extract its cross sectionnal dimensions. The red curves were used for dimensionning. They are fitted by eye on the micrograph's scale bar and the on the edges of the pillar.}
    \label{fig:s2}
\end{figure}

The cross-sectional dimensions of antennas were assessed using top down SEM as shown in Figure \ref{fig:s2}. By repeating the same measurement over 4 pillars in each array, statistics on these dimensions were obtained and are tabulated in Table\;1 of the main text. The uncertainties quoted include both those related to the fitting of the ellipse, inter-specimen variations and estimating the thickness of conductive material deposited. 

The antennas were fabricated on a non-conductive substrate. In order to obtain SEM micrographs, a thin gold layer was sputtered on the sample to minimize the presence of artefacts caused by charging effects. The thickness of this gold layer was measured to be $14\pm5$\;nm using ultraviolet light absorption\cite{kossoy_optical_2015}. The dimensions shown in Table\;1 have been corrected for this additional thickness.

\section{Supplementary DFS spectra}
\begin{figure}[h!]
    \centering
    \includegraphics{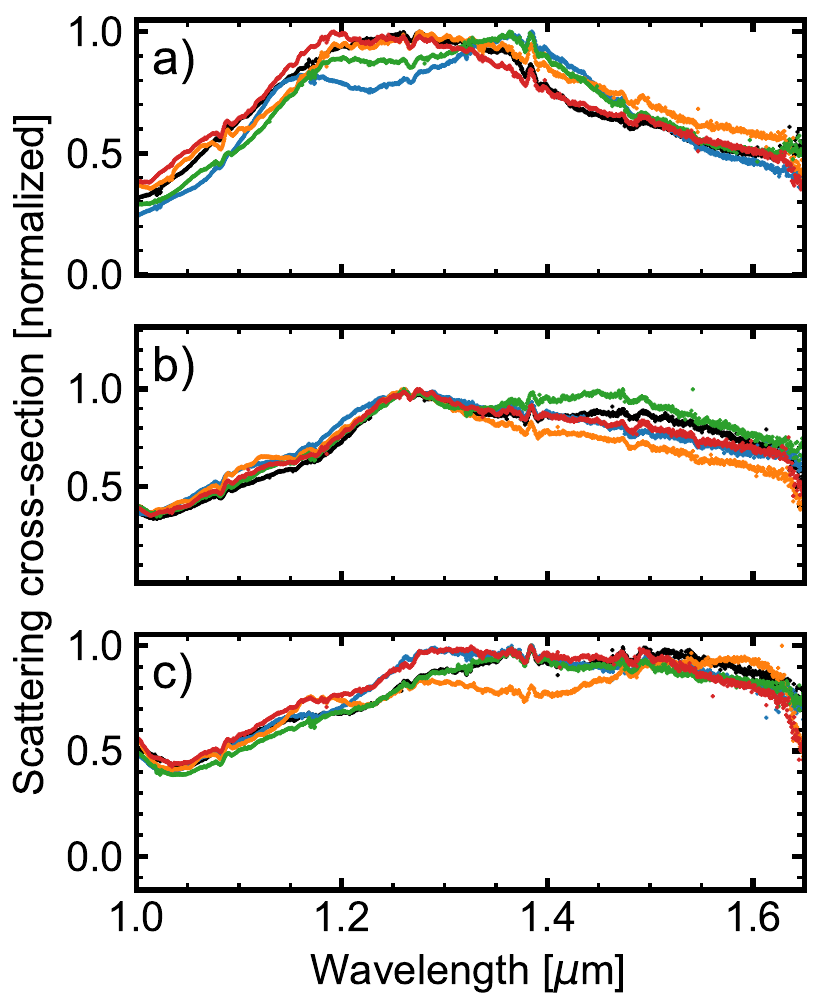}
    \caption{Dark field scattering spectra of pillars in arrays I, II and III (in a), b) and c) respectively). The black dots correspond to the spectra shown in Figure 2 of the main text.}
    \label{fig:adddfss}
\end{figure}

In the main text, the DFS spectrum of only one pillar per array was reported. In Figure \ref{fig:adddfss}, the DFS spectra of other pillars from the same array are plotted as well. Despite minor discrepancies in mode energy and scattering cross-section, the scattering spectra are similar.

\begin{figure}[h!]
    \centering
    \includegraphics[width=0.25\textwidth]{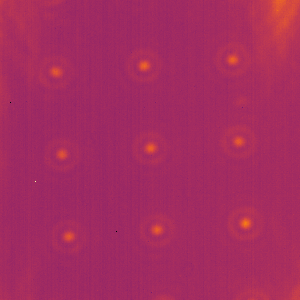}
    \caption{Typical infrared dark field scattering image of a pillar array. Individual pillars are clearly visible.}
    \label{fig:dfssimage}
\end{figure}

As can be seen from a typical dark field scattering image presented in Figure \ref{fig:dfssimage}, the collected signal is exclusively attributable to scattering of the incident light by the pillars.
During a DFS spectroscopy measurement, the slit is closed around a single column of pillars, vertically discriminating the spectra of single pillars.
\section{Supplementary instantaneous THG efficiency curves}
\begin{figure}[h!]
    \centering
    \includegraphics[width=\textwidth]{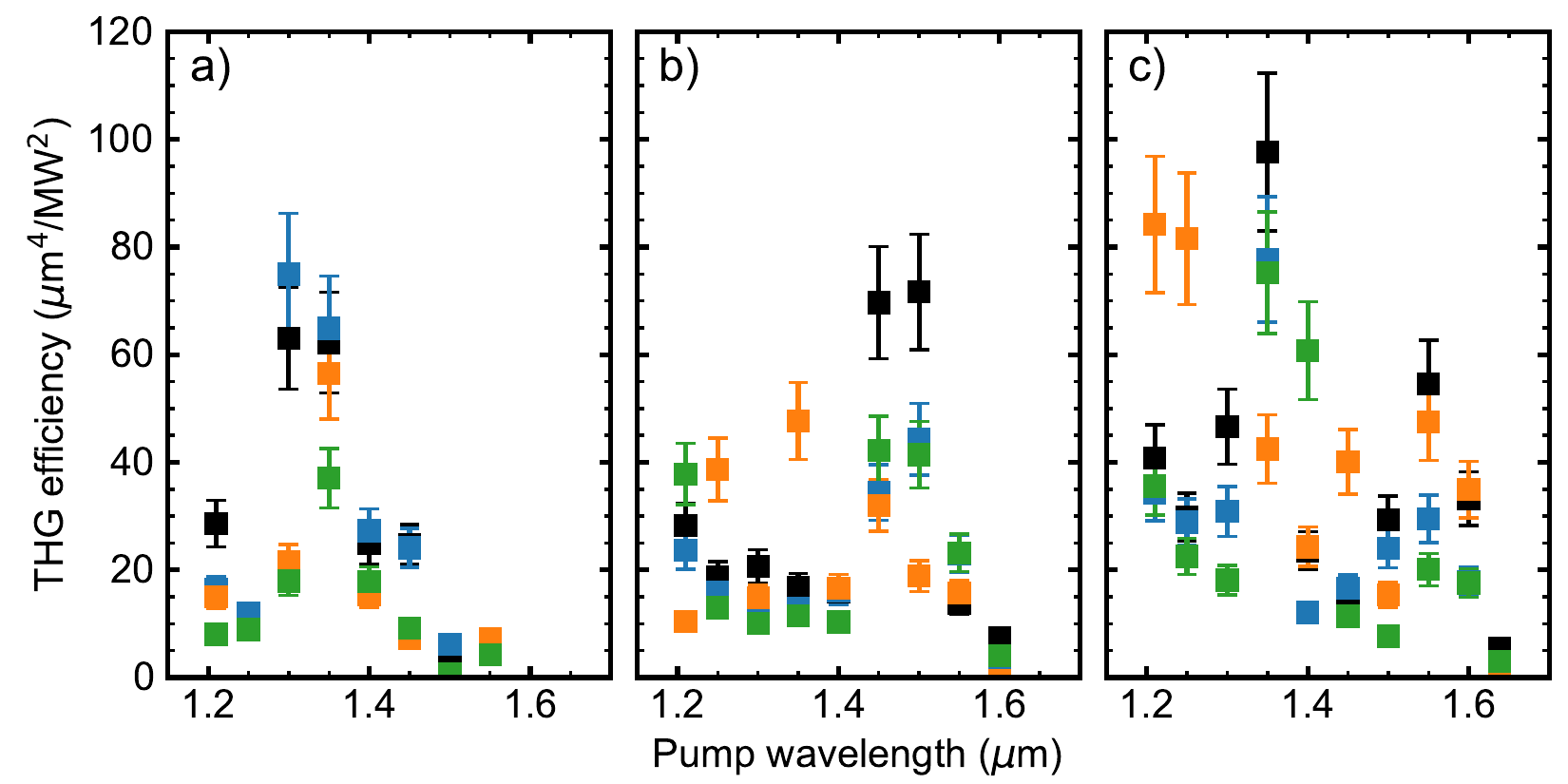}
    \caption{Instanteneous THG efficiency of pillars in arrays I, II and III (in a), b) and c) respectively). Each color corresponds to a different pillar. The black squares correspond to the data shown in Figure 5 of the main text.}
    \label{fig:thgwavedepadd}
\end{figure}

The instantaneous THG generation efficiencies measured for other pillars than those studied in the main text are plotted in Figure \ref{fig:thgwavedepadd}.

\section{Setup used for the optical characterisation}
\begin{figure}[h!]
    \centering
    \includegraphics[width=\textwidth]{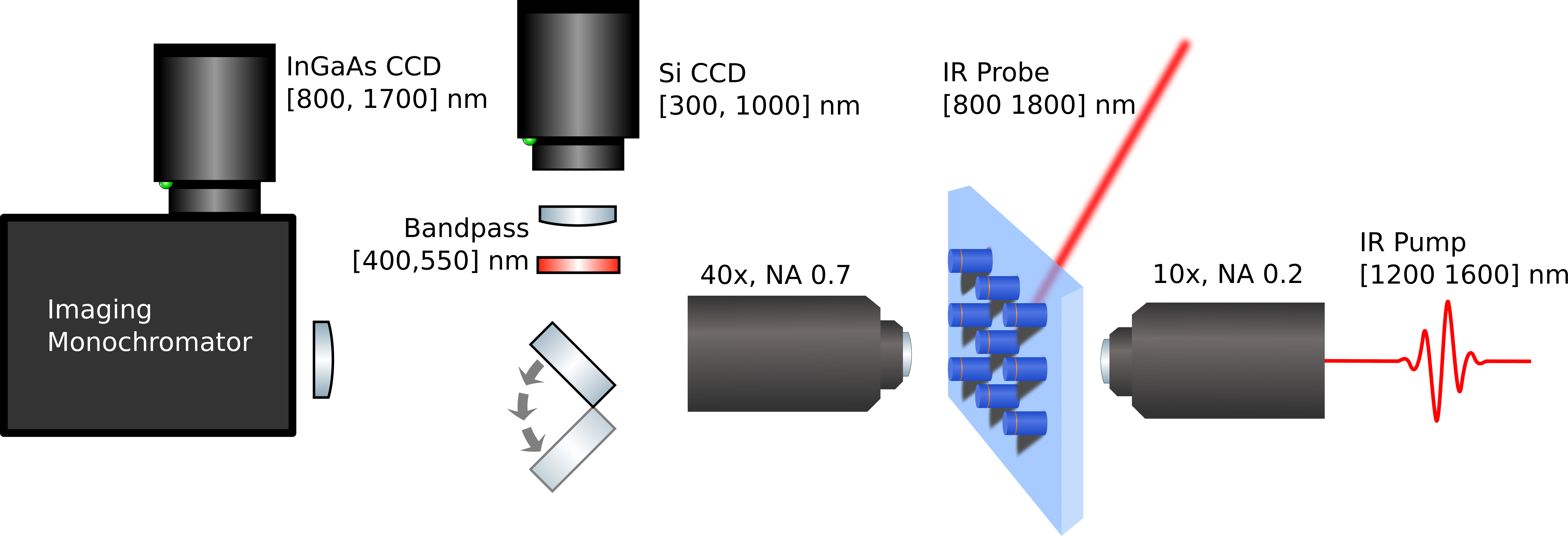}
    \caption{A diagram of the optical characterisation setup.}
    \label{fig:setup}
\end{figure}

A diagram of the experimental apparatus used to measure both dark field scattering spectra and third harmonic generation is presented in Figure\;\ref{fig:setup}.
This setup is described in detail in the Methods section of the main text.

\section{Simulations of the TH emission}
\begin{figure}[h!]
    \centering
    \includegraphics[width=\textwidth]{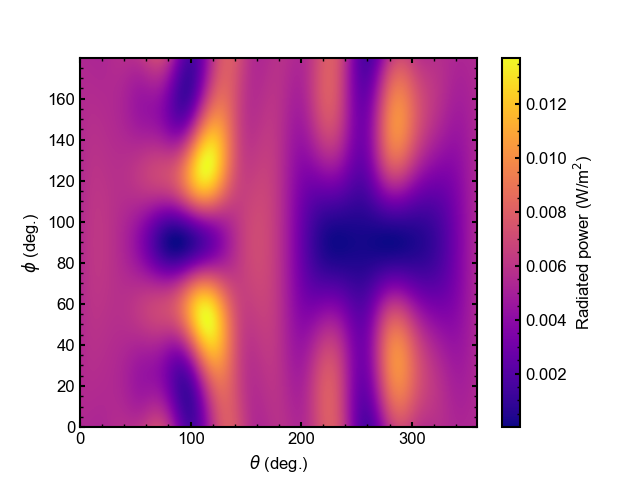}
    \caption{Radiation pattern from two out of phase dipoles placed inside the ITO slab of a pillar antenna.}
    \label{fig:radpat}
\end{figure}

We modelled the TH radiation pattern (shown in Figure\;\ref{fig:radpat}) and outcoupling efficiency by placing two 10\;nm-long out of phase dipoles at the two maxima of field enhancement in the ITO thin-film of a photonic gap antenna. This simulations was performed using a finite-difference time-domain method for a substrate-less model.
98\% of the TH is lost to the strong absorption of a-Si at this frequency.
Of the 2\% that couples out to the far-field, only 23\% are collected by our objective.

\bibliography{suppinfo}